\documentclass[%
 reprint,
 amsmath,amssymb,
 aps,
]{revtex4-2}
\usepackage{tikz-cd}
\usepackage{graphicx,amsmath,amssymb,algpseudocode,physics}
\usepackage{subcaption}
\usepackage{dcolumn}
\usepackage{bm}

\begin{document}

\preprint{APS/123-QED}

\title{Mixed-state extension of the holographic signal inequality}%

\author{Joydeep Naskar}
\affiliation{Department of Physics, Northeastern University, Boston, MA 02115, USA,}
\affiliation{Kavli Institute for Theoretical Physics, Santa Barbara, CA 93106, USA,}
\affiliation{Beijing Institute of Mathematical Sciences and Applications, Huaibei Town, Huairou District, Beijing 101408, China.}
\email{naskar.j@northeastern.edu}
\date{\today}

\begin{abstract}
A novel inequality involving the residual information and genuine multi-entropy was proposed in \cite{Balasubramanian:2025hxg} for tripartite holographic pure states, using which it was argued, that purely GHZ-like tripartite entanglement is not allowed in holography. In this work, we generalize this holographic signal inequality to mixed states. In a minimal extension, we compute the reflected genuine multi-entropy following \cite{Yuan:2024yfg} and find a class of holographic geometries that violate this minimally extended inequality due to a vanishing Markov gap. We can symmetrize this prescription, where instead of computing the residual information on the given mixed state $\rho_{ABC}$, we compute it on its canonical purification. The inequality is restored on the canonically purified state. Finally, we conjecture a new inequality for tripartite holographic states and give supporting evidence.
\end{abstract}

\maketitle


\section{Introduction}\label{sec:intro}
Quantum entanglement is a largely studied subject of the last century. We have made considerable progress towards understanding bipartite entanglement\cite{quantum-entanglement-review} between subsystems, and yet multipartite entanglement is not very well understood.

For holographic states, the holographic entropy cone (HEC) has provided a systematic framework to study constraints on \emph{bipartite} correlations between multiparty subregions on the boundary CFT (see \cite{Bao:2015bfa,Bao:2024azn,Bao:2025sjn,Avis:2021xnz,Czech:2021rxe,Hernandez-Cuenca:2022pst,Grimaldi:2026lbq,Hubeny:2025bjo,Hernandez-Cuenca:2023iqh,Hubeny:2025hst,Grimaldi:2025jad,Czech:2026tgj,Czech:2025jnw,Czech:2025tds,Czech:2024rco,Bao:2024vmy,Grado-White:2025jci,Zhao:2026smt,Bao:2024obe} for some recent samples). The holographic entropy cone for $n$ subregions comprises a finite number of \emph{tight} entropy inequalities that cannot be improved further. These inequalities hold for any holographic geometry, whose entanglement entropy (to leading order) is given by the Ryu-Takayanagi (RT) formula\cite{Ryu:2006ef}\footnote{Considering subleading corrections, if the bulk subregion entropies satisfy the HEC, the boundary subregion entropies also satisfy them \cite{Akers:2021lms}.}\footnote{It has been shown in \cite{Naskar:2024mzi,Bao:2015boa,Bousso:2025mdp} that these inequalities are satisfied beyond the scope of their original use.}. The RT formula says that for a boundary subregion $A$, the leading order entanglement entropy $S(A)$ is given by,
\begin{equation}\label{eq:RT-formula}
    S(A)= \frac{\text{area}(\gamma_A)}{4G_N},
\end{equation}
where $\gamma_A$ is the codimension-two minimal surface in the bulk --- equivalently, the codimension-one minimal surface in the bulk time slice --- homologous to the boundary subregion $A$, and $G_N$ is Newton's constant \footnote{Note that the HEC results derived using RT formula hold for the static scenario only. While no violation for their covariant generalization is known, with extensive evidence supporting in favor\cite{Grado-White:2024gtx,Grado-White:2025jci,Grimaldi:2025jad,Zhao:2026smt}, a rigorous proof for the same is absent.}.

There are some other approaches towards studying multipartite entanglement in holography, including L-entropy\cite{Basak:2024uwc}, multipartite entanglement signals\cite{Balasubramanian:2024ysu,Bao:2025psl,Balasubramanian:2025jhq,Ju:2026xfh,Gadde:2026msg,Ju:2025abf}, holographic duals to multi-entropy\cite{Gadde:2022cqi,Gadde:2023zzj,Gadde:2024taa,Iizuka:2025elr}, upper-bounding entropy relations \cite{Ju:2025tgg,Ju:2024kuc} etc. However, there is no general consensus on a universally accepted measure for multipartite entanglement, so for time being, we will content ourselves with their diagnostic \emph{signals}.

The organization of this short work is as follows. We will review the relevant concepts and introduce the signal inequality in section \ref{sec:review}, and propose its mixed-state generalizations in section \ref{sec:mixed}. In the \emph{minimal extension}, we show that certain geometries can violate this generalized signal inequality. However, we also give a \emph{symmetric extension} and provide evidence in its support. Next, we conjecture a new candidate inequality for tripartite correlation signals in section \ref{sec:conj}. We conclude with some discussions and future directions.

\section{A Brief Review of the Signal Inequality}\label{sec:review}
Let us first define the Renyi multi-entropy. Instead of giving a general definition, we will specialize to the tripartite case, which is relevant for the discussion that follows. Consider a tripartite pure state $\ket{\psi}$ on the Hilbert space $\mathcal{H}=\mathcal{H}_A\otimes\mathcal{H}_B\otimes\mathcal{H}_C$, 
\begin{equation}\label{eq:state-psi}
    \ket{\psi}=\sum_{i_A=1}^{d_A}\sum_{i_B=1}^{d_B}\sum_{i_C=1}^{d_C}\psi_{i_A i_B i_C} \ket{e_A^{i_A}}\ket{e_B^{i_B}}\ket{e_C^{i_C}},
\end{equation}
where $\{\ket{e^{i_X}_X}\}$ is a basis of $\mathcal{H}_X$ and $d_X$ is its dimension.

The tripartite Renyi multi-entropy is defined as follows \cite{Iizuka:2025ioc},
\begin{equation}\label{eq:renyi-3-multi}
    S^{(3)}_n(A:B:C)=\frac{1}{1-n} \frac{1}{n} \log{\frac{Z_n^{(3)}}{(Z_1^{(3)})^{n^2}}},
\end{equation}
and $Z_n^{(3)}$ is given by,
\begin{equation}
    Z_n^{(3)}:=\bra{\psi}^{\otimes n^2}\Sigma_A(g_A) \Sigma_B(g_B) \Sigma_C(g_C) \ket{\psi}^{\otimes n^2},
\end{equation}
where $\Sigma_{A,B,C}$ are the twist operators corresponding to the action of the permutation group elements $g_{A,B,C}$ on the replica indices. We are not explicitly writing down the permutation elements here, however, we state an example, e.g., for $n=2$, we have \cite{Harper:2025uui}
\begin{align}
    & S_2^{(3)}(A:B:C)= -\frac{1}{2} \log{\mathcal{\varepsilon}},\\
    \mathbf{\varepsilon}=&\psi_{iwa}\psi_{jxb}\psi_{kyc}\psi_{lzd}\bar{\psi}^{jya}\bar{\psi}^{izb}\bar{\psi}^{lwc}\bar{\psi}^{kxd},
\end{align}
where we have chosen the normalization $Z_1^{(3)}=1$. The replica construction behind these definitions, and the explicit permutation elements $g_{A,B,C}$, are collected in appendix \ref{app:multi-entropy}.
The multi-entropy $S^{(3)}(A:B:C)$ is defined by taking the $n\rightarrow 1$ limit,
\begin{equation}\label{eq:multi-n1}
    S^{(3)}(A:B:C)= \lim_{n\rightarrow 1} S^{(3)}_n(A:B:C).
\end{equation}

An important quantity of interest that we will discuss now is the \emph{genuine multi-entropy}\cite{Iizuka:2025ioc,Iizuka:2025caq}. It is built using the multi-entropy. More precisely, the genuine Renyi multi-entropy $GM^{(3)}_n$ for tripartite states is defined as,
\begin{equation}
\begin{aligned}\label{eq:GME-def1}
    & GM^{(3)}_n(A:B:C) \\ & = S^{(3)}_n(A:B:C) \\
    & -\frac{1}{2}\left(S^{(2)}_n(AB:C)+S^{(2)}_n(AC:B)+S^{(2)}_n(BC:A) \right)
\end{aligned}
\end{equation}
where $n$ is the Renyi index, and $S^{(q)}_{n}$ is the $q$-party Renyi multi-entropy. Note that we have defined $S^{(3)}_{n}$ in eq. \ref{eq:renyi-3-multi} and $S^{(2)}_{n}$ is ordinary Renyi entropy capturing bipartite correlation. Appendix \ref{app:genuine} explains why this particular combination is the unique one that removes all bipartite contributions. We define the genuine multi-entropy $GM^{(3)}(A:B:C)$ by taking the $n\rightarrow 1$ limit,
\begin{equation}\label{eq:GME}
    GM^{(3)}(A:B:C)= \lim_{n\rightarrow 1} GM^{(3)}_n(A:B:C).
\end{equation}

\begin{figure}[h!]
		\centering
		\includegraphics[width=0.35\textwidth]{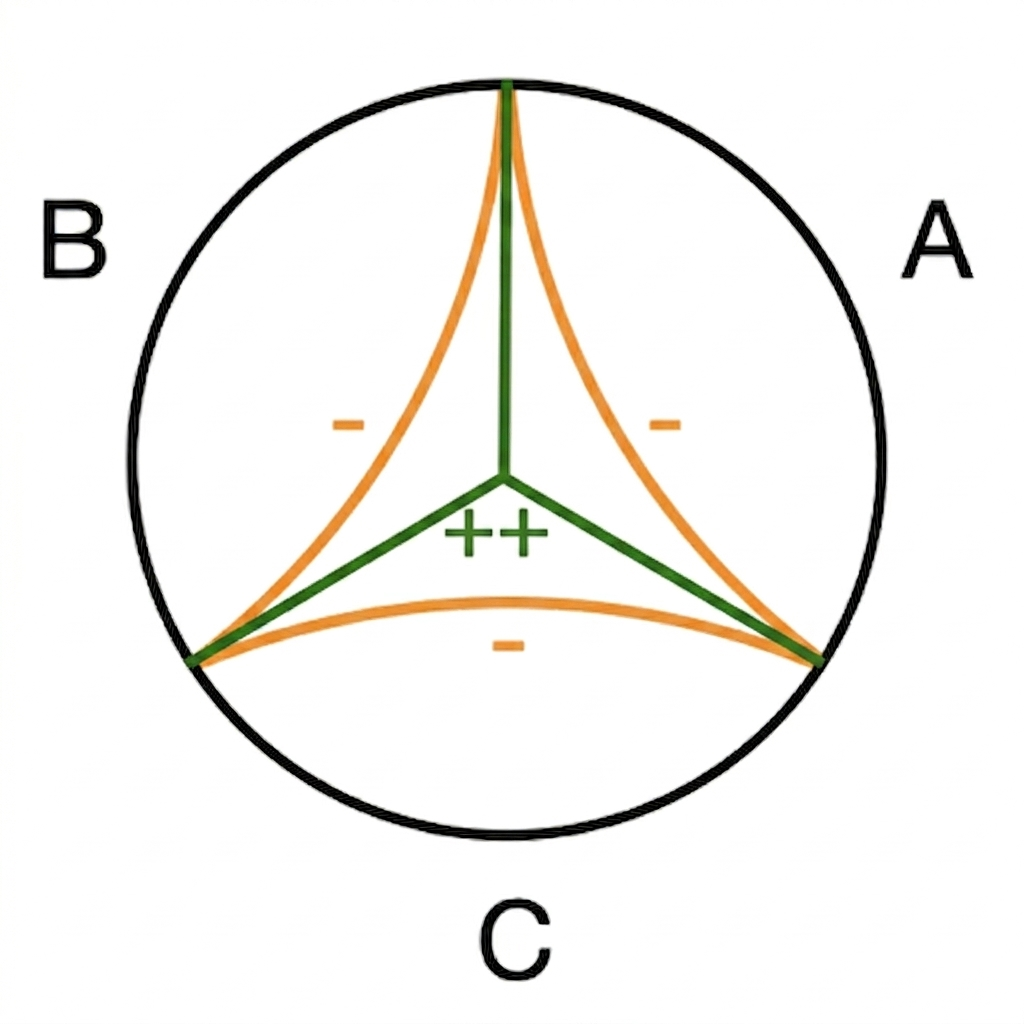}
	\caption{The green curve gives $S^{(3)}(A:B:C)$ and three orange curves give $S^{(2)}(AB:C), S^{(2)}(AC:B), S^{(2)}(BC:A)$ respectively. It is easy to see from the geometry that $GM^{(3)}(A:B:C)$ defined in eq. \ref{eq:GME} is positive definite, where the lengths of the orange curves are subtracted from twice the length of the green curve.}
	\label{fig:GM3}
\end{figure}

Now let us talk about holography. Consider the simplest case of a Cauchy slice in $AdS_3/CFT_2$. The holographic dual of the quantity $S^{(3)}(A:B:C)$ (shown in figure \ref{fig:GM3}) is given by,
\begin{equation}\label{eq:multi-holo}
    S^{(3)}(A:B:C)= \frac{1}{4G_N}\min_{O} \left(|\gamma_{AO}|+|\gamma_{BO}|+|\gamma_{CO}|\right),
\end{equation}
where $O$ is the trijunction point in the bulk and $|\gamma_{OX}|$ is the bulk-to-boundary geodesic between bulk point $O$ and the first end point (clockwise) of the boundary subregion $X$. Thus, we have to define this geodesic length with a short-distance cutoff $\epsilon$. Moreover, we have $S^{(2)}(AB:C)=S(C)$ and likewise we have $S(A)$ and $S(B)$, which are entanglement entropies, whose holographic duals are minimal surfaces given by the RT formula \eqref{eq:RT-formula}. Therefore, $GM^{(3)}(A:B:C)$ can be geometrically computed (see figure \ref{fig:GM3}). An interesting fact about $GM^{(3)}(A:B:C)$ is that it is \emph{finite} and cutoff-independent, which is lower bounded by \cite{Iizuka:2025caq}
\begin{equation}\label{eq:GME_AdS3}
    GM^{(3)}(A:B:C)\geq\frac{3}{4G_N}\log{\frac{2}{\sqrt{3}}}.
\end{equation}

\begin{figure}[h!]
		\centering
		\includegraphics[width=0.35\textwidth]{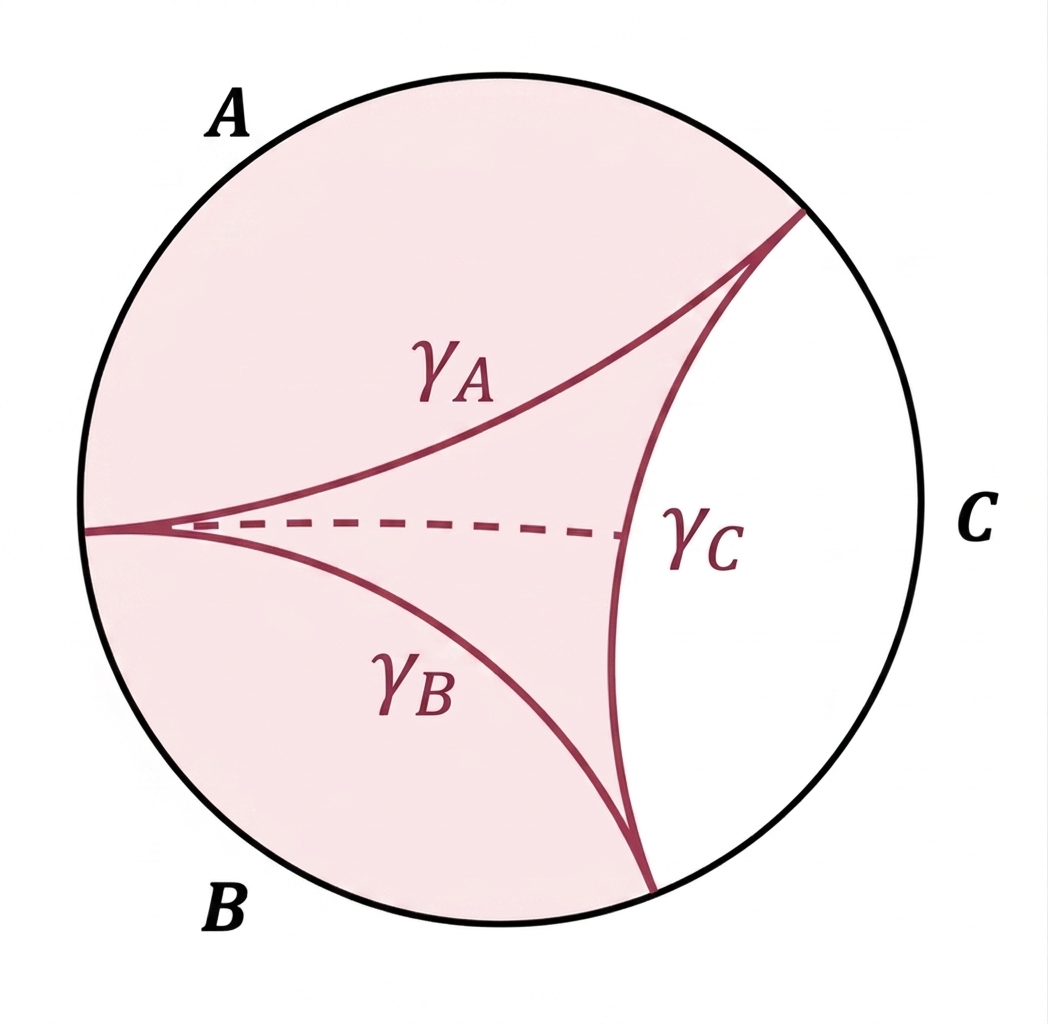}
	\caption{In this figure, the maroon curves $\gamma_A$, $\gamma_B$ and $\gamma_C$ are the minimal surfaces giving $S(A), S(B), S(AB)$ respectively. The dashed maroon curve is the entanglement wedge cross-section $\Gamma_{AB}$ of the entanglement wedge of $AB$.}
	\label{fig:Markov_gap}
\end{figure}

The second quantity of interest is the Markov gap \cite{Hayden:2021gno} (also called residual information) $R^{(3)}(A:B)$, defined as the difference of the reflected entropy $S_R(A:B)$ and the mutual information $I(A:B)$ of a reduced state $\rho_{AB}$ of two parties,
\begin{equation}\label{eq:Markov_gap}
    R^{(3)}(A:B)=S_R(A:B)-I(A:B),
\end{equation}
where $I(A:B)=S(A)+S(B)-S(AB)$. The Markov gap has several interesting properties. However, we will just focus on its holographic dual shown in figure \ref{fig:Markov_gap}, i.e., for a pure state on the full boundary CFT, the Markov gap is given by,
\begin{equation}
     R^{(3)}(A:B)= 2\Gamma_{AB} - \left(\gamma_{A}+\gamma_{B}-\gamma_{C}\right),
\end{equation}
where $\Gamma_{AB}$ is the entanglement wedge-cross section of the entanglement wedge of $AB$.

In a recent work \cite{Balasubramanian:2025hxg}, an inequality was proposed for tripartite states in holography,
\begin{equation}\label{eq:signal_ineq_pure}
    \frac{1}{2} R^{(3)}(A:B) \geq  GM^{(3)}(A:B:C),
\end{equation}
which we refer to as \emph{holographic signal inequality}(HSI) and is shown to hold for pure tripartite states on the boundary CFT. Note that in the presentation of inequality \ref{eq:signal_ineq_pure}, we have exercised the choice of tracing over $C$ and working with $\rho_{AB}$. There are two other inequalities analogous to eq \ref{eq:signal_ineq_pure} involving $\rho_{BC}$ and $\rho_{AC}$ respectively, all three sharing the same right-hand side, since $GM^{(3)}(A:B:C)$ is permutation symmetric. For simplicity of presentation, we will stick to the inequality \ref{eq:signal_ineq_pure} involving $\rho_{AB}$. This inequality led to an important conclusion that purely GHZ-like tripartite entanglement is ruled out in holography. This is because the Markov gap is zero for the GHZ and generalized-GHZ states. Put differently, the HSI says that no holographic tripartite pure state can simultaneously have zero Markov gap and positive genuine multi-entropy.

The generalization of the HSI for mixed states was not discussed in \cite{Balasubramanian:2025hxg}, so we explore this question.

\section{Holographic signal inequality for mixed states}\label{sec:mixed}

\subsection{A minimal extension}

The signal inequality can be generalized for mixed states as follows. The Markov gap $R^{(3)}(A:B)$ is defined just the same as in eq. \eqref{eq:Markov_gap} as the difference of reflected entropy $S_R(A:B)$ and mutual information $I(A:B)$. Clearly, the holographic interpretation of $S_R(A:B)$ remains unchanged as \emph{twice} the cross-section of the entanglement wedge of $AB$, i.e. $S_R(A:B)=2\Gamma_{AB}$ \cite{Dutta:2019gen}, since $S_R(A:B)$ is computed using the density matrix $\rho_{AB}$ supported on two subregions $A$ and $B$.

We can generalize the multi-entropy to its mixed-state version called \emph{reflected multi-entropy}\cite{Yuan:2024yfg} by canonical purification. Let us walk through the formulation of reflected multi-entropy. Consider a pure state $\ket{\psi_{AcBaCb}} \in \mathcal{H}_{AcBaCb}$, where $a,b,c$ label auxiliary purifying systems that are traced out to obtain the mixed state $\rho_{ABC}$. The mixed state $\rho_{ABC}$ can be canonically purified,
\begin{equation}
\begin{aligned}\label{eq:rho_can_pur}
\ket{\sqrt{\rho_{ABC}}}& =\ket{\sqrt{\text{Tr}_{abc}\ket{\psi_{AcBaCb}}\bra{\psi_{AcBaCb}}}}\\
& \in (\mathcal{H}_A\otimes\mathcal{H}^*_A)\otimes (\mathcal{H}_B\otimes\mathcal{H}^*_B)\otimes (\mathcal{H}_C\otimes\mathcal{H}^*_C),
\end{aligned}
\end{equation}
then, the reflected multi-entropy is defined as,
\begin{equation}\label{eq:mixed-state-multi}
    S^{(3)}_R(A:B:C)=S^{(3)}(AA^*:BB^*:CC^*)_{\ket{\sqrt{\rho_{ABC}}}},
\end{equation}
where the subscript in the RHS is a reminder that we are working with the canonically purified state $\ket{\sqrt{\rho_{ABC}}}$. 

For the bipartite correlations, recall that we were computing quantities of the type $S^{(2)}(AB:C)$ and now they are not simply given by $S(C)$. Instead, one needs to modify it to
\begin{equation}
    S^{(2)}_R(AB:C)=S^{(2)}(ABA^*B^*:CC^*)_{\ket{\sqrt{\rho_{ABC}}}},
\end{equation}
for the canonically purified state $\ket{\sqrt{\rho_{ABC}}}$.

\begin{figure}[h!]
		\centering
		\includegraphics[width=0.3\textwidth]{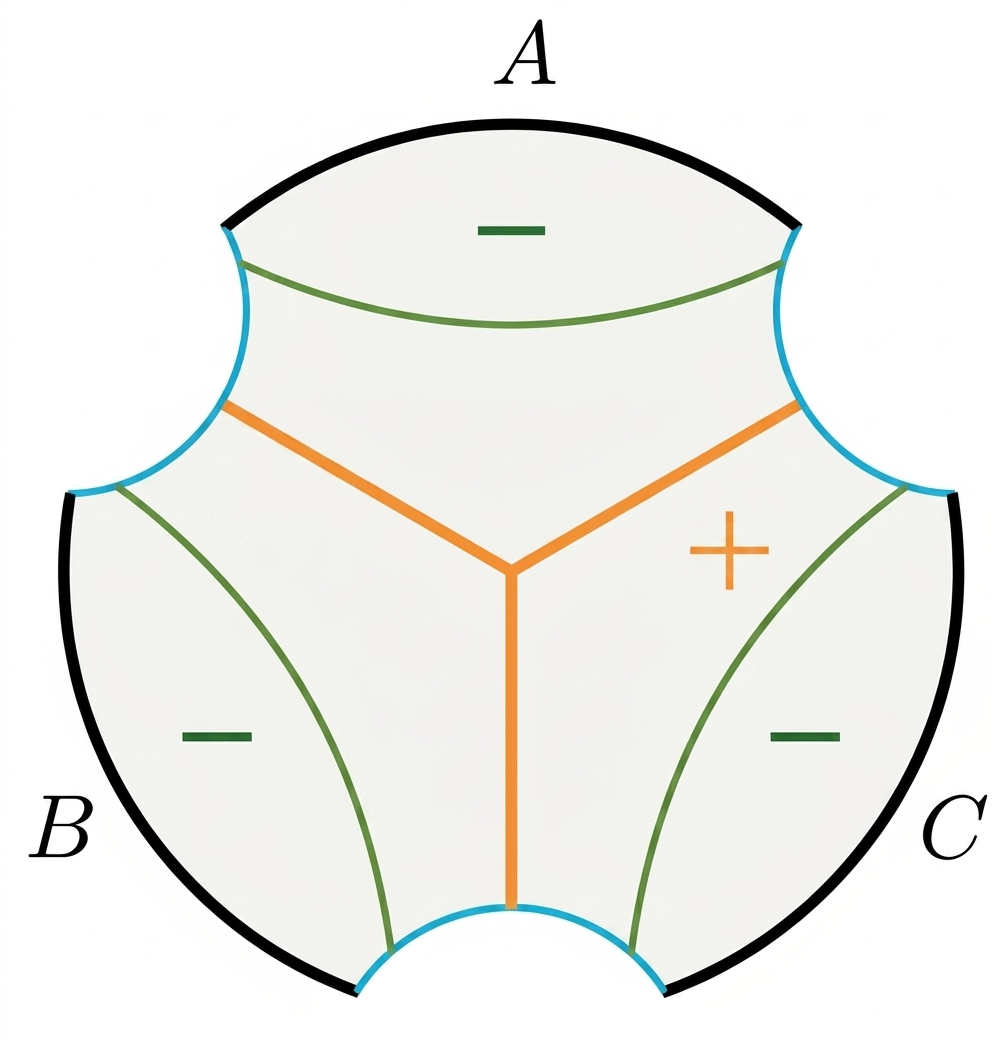}
	\caption{The orange curve gives $\frac{1}{2}S^{(3)}_R(A:B:C)$ and the three green curves give $\frac{1}{2}S^{(2)}_R(AB:C), \frac{1}{2}S^{(2)}_R(AC:B), \frac{1}{2}S^{(2)}_R(BC:A)$ respectively. Again, it is easy to see from the geometry that the $GM^{(3)}_R(A:B:C)$ defined in eq \ref{eq:reflected-GM} is positive definite.}
	\label{fig:GM3_reflected}
\end{figure}

We can construct the reflected multi-entropies $S^{(2)}_R(AC:B)$ and $S^{(2)}_R(BC:A)$ similarly. The proposed holographic duals of reflected multi-entropies $S^{(2)}_R(AB:C)$ and $S^{(3)}_R(A:B:C)$ (and others) are shown in figure \ref{fig:GM3_reflected}. This construction ensures that $S^{(2)}_R(AB:C)$ is cutoff-independent and finite.

The definition of the $GM^{(3)}_R$ remains the same as before, except now using the reflected multi-entropy $S^{(q)}_R$. Explicitly, we have
\begin{equation}\label{eq:reflected-GM}
\begin{aligned}
    &GM^{(3)}_R(A:B:C)\\&=\frac{1}{2}S^{(3)}_R(A:B:C)\\
    &-\frac{1}{4}\left(S^{(2)}_R(AB:C)+S^{(2)}_R(AC:B)+S^{(2)}_R(BC:A) \right),
\end{aligned}
\end{equation}
where the overall factor of $\frac{1}{2}$ is to account for the double-copy. It is useful to note that in the limit when the gaps between boundary subregions $A,B,C$ shrink to zero, such that $\rho_{ABC}$ approaches the global pure state $\ket{\psi}$, then $GM^{(3)}_R(A:B:C)$ approaches $GM^{(3)}(A:B:C)$. In fact, the constituent reflected quantities $S^{(3)}_R(A:B:C)$ and $S^{(2)}_R(AB:C)$ approach $2S^{(3)}(A:B:C)$ and $2S^{(2)}(AB:C)$ respectively.

So, now we can analogously define our signal inequality for mixed states as follows (analogous to eq \ref{eq:signal_ineq_pure}),
\begin{equation}\label{eq:signal_ineq_mixed}
    \frac{1}{2} R^{(3)}(A:B) \geq GM_R^{(3)}(A:B:C).
\end{equation}

\begin{figure}[h!]
		\centering
		\includegraphics[width=0.35\textwidth]{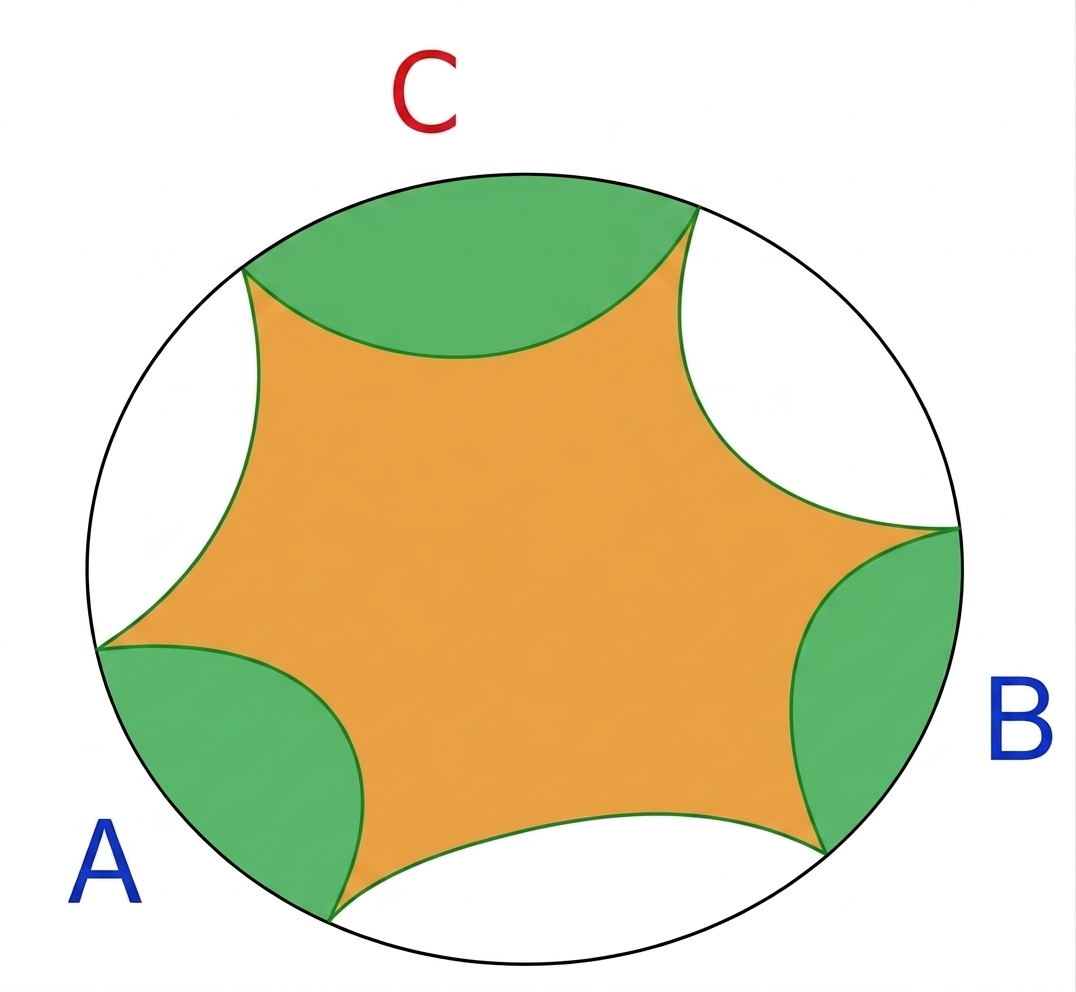}
	\caption{A schematic geometry in which the boundary subregions $A$, $B$ and $C$ are chosen symmetrically. Each of the three regions shaded in green is the entanglement wedge of a single subregion, so the entanglement wedge of every pair --- $AB$, $BC$ and $AC$ --- is disconnected, while the entanglement wedge of $ABC$ is connected, comprising the three green regions together with the orange region. This geometry therefore violates all three instances of the inequality eq \ref{eq:signal_ineq_mixed}, but satisfies the conjectured inequalities eq \ref{eq:new_conjecture} and eq \ref{eq:new_conjecture_red}.}
	\label{fig:example_geometry_violate}
\end{figure}

It is easy to see that inequality (\ref{eq:signal_ineq_mixed}) can be violated by holographic geometries. For example, consider a boundary configuration of subregions $A, B, C$, such that the entanglement wedge of $AB$ is disconnected, but that of $ABC$ is connected (for example, see figure \ref{fig:example_geometry_violate}). In this case, since the entanglement wedge of $AB$ is disconnected, both the entanglement wedge cross-section and the mutual information are identically zero, giving $\frac{1}{2} R^{(3)}(A:B)=0$. However, $GM^{(3)}_R(A:B:C)>0$, thus violating the inequality (\ref{eq:signal_ineq_mixed}). In the example considered in figure \ref{fig:example_geometry_violate}, since the subregions $A$, $B$ and $C$ are symmetric, the violation of HSI occurs for all three instances involving $\rho_{AB}$, $\rho_{BC}$ and $\rho_{AC}$ respectively. However, for an asymmetric choice of subregions, there is a possibility of a \emph{weaker} violation, where only a subset of the three HSIs are violated. Both sides of (\ref{eq:signal_ineq_mixed}) are evaluated in closed form for this configuration in appendix \ref{app:explicit}, which makes the range of subregion sizes over which the violation occurs explicit.

One technical argument against the minimal extension is that,  after canonical purification, the LHS and RHS are computed  on different states. While this is technically correct, recall that the original pure-state inequality (eq.~\ref{eq:signal_ineq_pure}) already has this feature, that the LHS Markov gap is defined via the canonical purification of $\rho_{AB}$ and the LHS mutual information is computed on the mixed state $\rho_{AB}$, while the RHS multi-entropy is defined on the global pure state $\ket{\psi} \in \mathcal{H}_A \otimes \mathcal{H}_B \otimes \mathcal{H}_C$. Our generalization lifts this construction one level higher by purifying the mixed state on the RHS as well, i.e., $R^{(3)}(A:B)$ continues to probe $\rho_{AB}$ via its canonical purification, while $GM^{(3)}_R$ probes $\rho_{ABC}$ via \emph{its} canonical purification. Moreover, it has the correct limiting behaviour when the mixed state $\rho_{ABC}$ approaches the global pure state $\ket{\psi}$. Furthermore, the lower bound for $GM^{(3)}_R(A:B:C)$ in the connected phase is given by,
\begin{equation}
    GM^{(3)}_R(A:B:C) \geq \frac{3}{4G_N}\log{\frac{2}{\sqrt{3}}},
\end{equation}
which follows from \cite{Yuan:2024yfg}. For the disconnected phase, where the regions $A,B,C$ are small enough and far apart, we have $GM^{(3)}_R(A:B:C)=0$, in the absence of genuine tripartite entanglement.

We emphasize that the minimal extension considered here is a natural choice based on standard techniques to compute Markov gap and reflected entropy for mixed states. As long as the Markov gap of the mixed state $\rho_{AB}$ is zero, any purification $\ket{\psi}_p$ of $\rho_{ABC}$ that is holographic will violate the HSI, provided the purifying system is grouped with one of the three parties so that $\ket{\psi}_p$ is genuinely tripartite and $GM^{(3)}$ is defined on it; the canonical purification (\ref{eq:rho_can_pur}) does this symmetrically, distributing the purifier as $AA^{*}|BB^{*}|CC^{*}$, and then $GM^{(3)}|_{\ket{\psi}_p}>0$ \cite{Iizuka:2025caq}. One may argue that we can have a holographic mixed state $\rho_{ABC}$ with its purification chosen such that $\ket{\psi}_{p}$ is not holographic. While technically this is possible, such a choice needs to be motivated. Moreover, while there is no proof of $GM^{(3)}\geq0$ for generic quantum states, there is increasing evidence for it, e.g., see \cite{Berthiere:2025toi}. More generally, we expect $GM^{(3)}(A:B:C)$ to be positive for a density matrix $\rho_{ABC}$ containing tripartite correlation, so the HSI will inevitably be violated by geometries like \ref{fig:example_geometry_violate} when the Markov gap is zero.

\subsection{A symmetric extension}
In this construction, we will first canonically purify the mixed state $\rho_{ABC}$ to $\ket{\sqrt{\rho_{ABC}}}$ defined in eq (\ref{eq:rho_can_pur}), and treat this as a pure state problem on the canonically purified state $\ket{\sqrt{\rho_{ABC}}}$. As seen before, the root cause of the violation was the vanishing Markov gap of the mixed state $\rho_{AB}$ and now we don't have to deal with $\rho_{AB}$. We will refer to this generalization of the HSI as the \emph{symmetric extension}.

With the pure state $\ket{\sqrt{\rho_{ABC}}}$ in hand, we will now trace over $C,C^*$ to get a mixed state $\hat{\rho}_{AB}^{A^*B^*}$, which is different from $\rho_{AB}$. We will compute the mutual information,
\begin{equation}\label{eq:sym_MI}
    I(AA^*:BB^*)=S(AA^*)+S(BB^*)-S(AA^*BB^*),
\end{equation}
which is now non-vanishing for the geometry in fig \ref{fig:example_geometry_violate}. To compute the reflected entropy $S_R(AA^*:BB^*)$, we will canonically purify the mixed state $\hat{\rho}_{AB}^{A^*B^*}$ to
\begin{equation}
\begin{aligned}
        &\ket{\sqrt{\hat{\rho}_{AB}^{A^*B^*}}}\\&\in (\mathcal{H}_{A_1}\otimes\mathcal{H}^*_{A_1})\otimes (\mathcal{H}_{A_2}\otimes\mathcal{H}^*_{A_2}) \\
        & \otimes(\mathcal{H}_{B_1}\otimes\mathcal{H}^*_{B_1})\otimes (\mathcal{H}_{B_2}\otimes\mathcal{H}^*_{B_2}), 
\end{aligned}
\end{equation}
which is different from the canonical purification of $\rho_{AB}$. Now we can write the symmetric extension of the HSI as,
\begin{equation}\label{eq:sym-ext-HSI-double}
    \frac{1}{2}{R}^{(3)}(AA^*:BB^*) \geq GM^{(3)}(AA^*:BB^*:CC^*),
\end{equation}
whose single-copy version we propose as,
\begin{equation}\label{eq:sym-ext-HSI-single}
    \frac{1}{2}R_R^{(3)}(A:B;[C]) \geq GM_R^{(3)}(A:B:C),
\end{equation}
where we call ${R}_R^{(3)}(A:B;[C])$ as the \emph{C-excised reflected residual entropy}, defined as,
\begin{equation}\label{eq:reflected-residual-entropy}
    {R}_R^{(3)}(A:B;[C])=\frac{1}{2}\left(S_{RR}(A:B;[C])-I_R(A:B;[C]) \right),
\end{equation}
where $S_{RR}(A:B;[C])$ and $I_R(A:B;[C])$ are the C-excised reflected entropy and C-excised mutual information respectively; both are constructed in appendix \ref{app:c-excised}.

\begin{figure}[h!]
		\centering
		\includegraphics[width=0.35\textwidth]{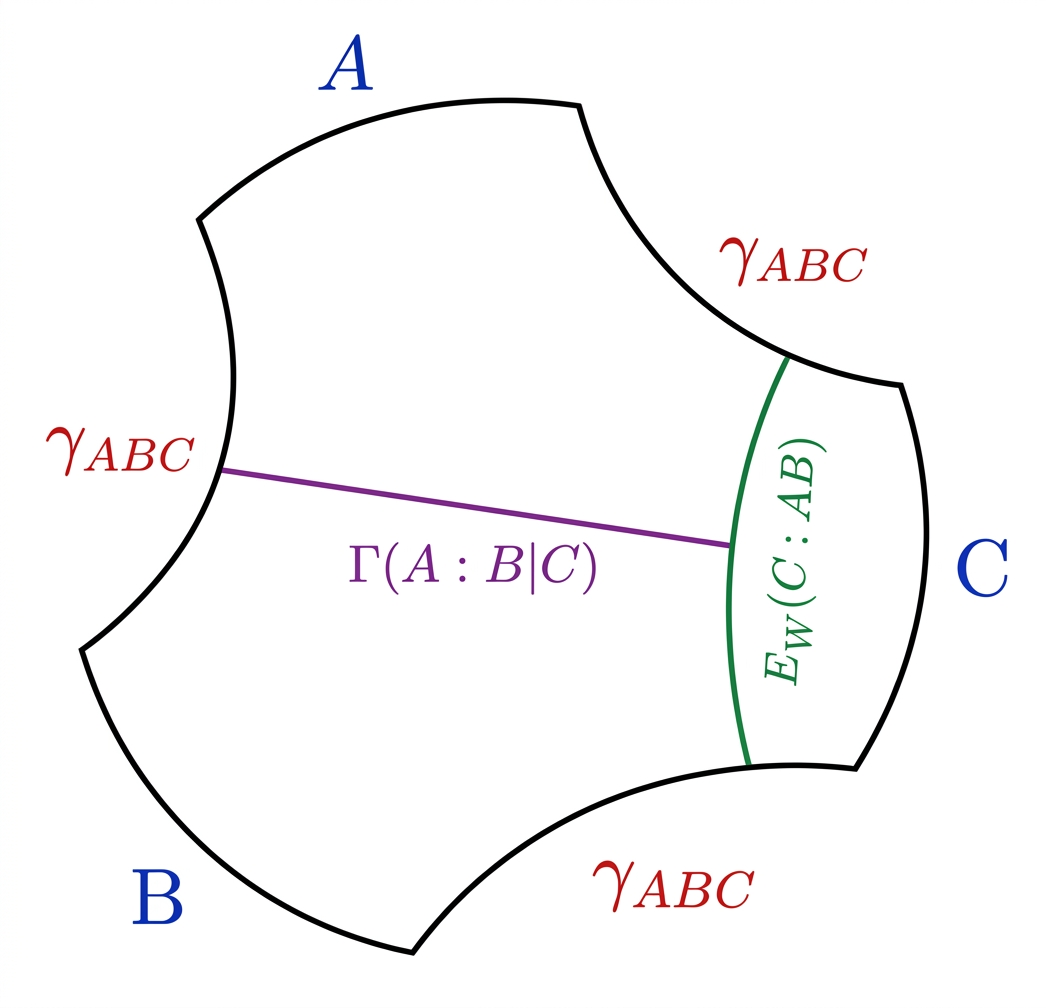}
	\caption{The cross-section of the $C$-excised connected entanglement wedge of $AA^*BB^*$, projected onto a single copy of the geometry, is denoted $\Gamma(A:B|C)$; the cross-section in the doubled geometry consists of two mirror copies of it, so that $S_{RR}(A:B;[C])=4\,\Gamma(A:B|C)$ as derived in appendix \ref{app:c-excised}. The entanglement wedge of $AB$ however continues to remain disconnected.}
    \label{eq:C-excised-reflected}
\end{figure}

The geometric interpretation of the reflected entropy $S_R(AA^*:BB^*)$ after projecting on the single copy of the geometry is proposed in figure \ref{eq:C-excised-reflected}.

As evident from figure \ref{eq:C-excised-reflected}, in the limit when the boundary state is a pure state, the inequality \ref{eq:sym-ext-HSI-single} approaches the HSI for pure states. For generic mixed states, if the canonically purified states are holographic, then the burden of proof of inequality \ref{eq:sym-ext-HSI-double} falls upon \cite{Balasubramanian:2025hxg}, since it is precisely the pure-state HSI applied to $\ket{\sqrt{\rho_{ABC}}}$. The single-copy form \ref{eq:sym-ext-HSI-single} then follows from the identifications $GM^{(3)}(AA^*:BB^*:CC^*)=2\,GM^{(3)}_R(A:B:C)$ and ${R}^{(3)}(AA^*:BB^*)=2\,{R}_R^{(3)}(A:B;[C])$, under which both forms reduce to ${R}_R^{(3)}(A:B;[C])\geq 2\,GM^{(3)}_R(A:B:C)$. For the class of geometries represented by fig \ref{fig:example_geometry_violate}, the residual information is able to capture the tripartite correlations between $AA^*$ and $BB^*$ in the state $\hat{\rho}_{AB}^{A^*B^*}$ that it failed to capture between $A$ and $B$ in the state $\rho_{AB}$. So the inequality is satisfied by the geometry in fig \ref{fig:example_geometry_violate}.

A skeptical reader may complain that we have changed the problem by changing the state $\rho_{ABC}$ to its canonical purification, which fits right in the paradigm of \cite{Balasubramanian:2025hxg}. While it is a fair criticism, we would like to think that this is an appropriate way to deal with mixed states for computing the residual information to faithfully capture the correlations of $ABC$.

\section{A New Conjecture Inequality for Holographic States}\label{sec:conj}
\begin{figure}[h!]
		\centering
		\includegraphics[width=0.35\textwidth]{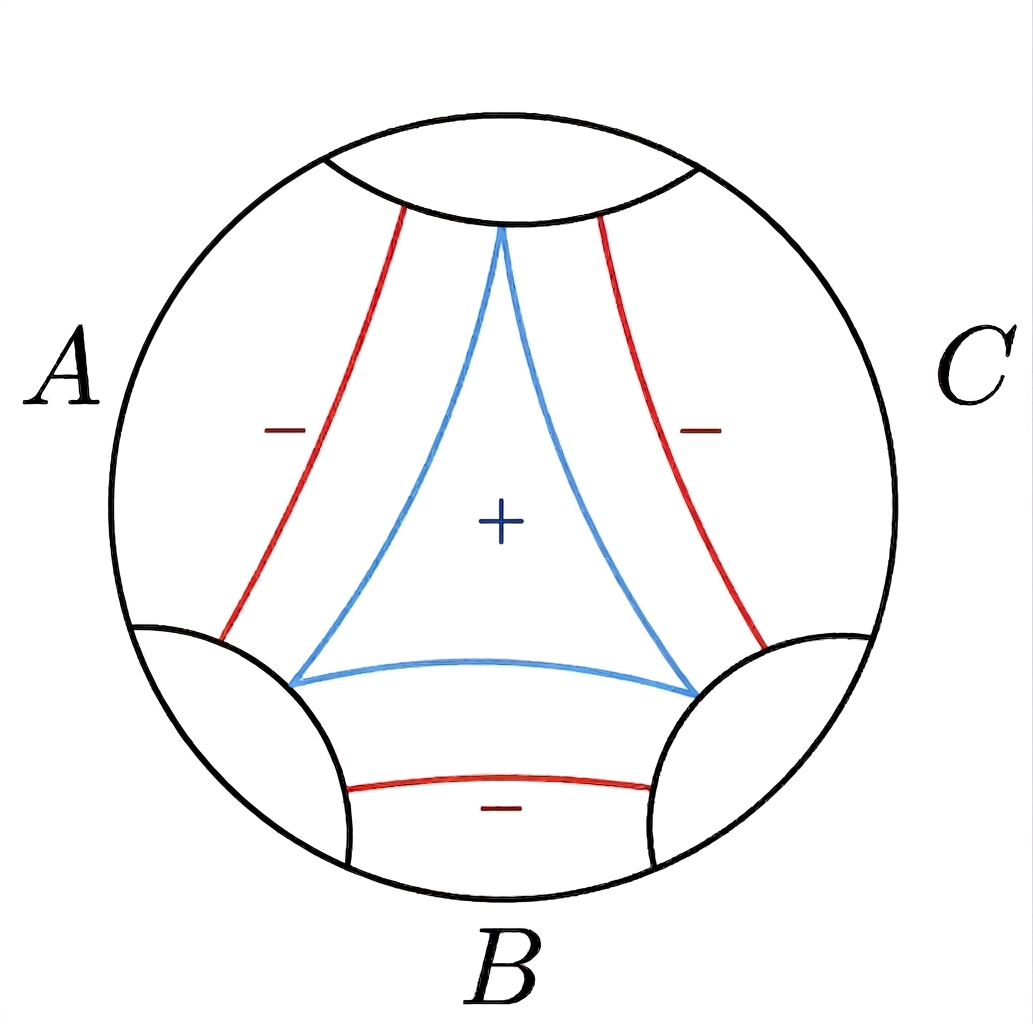}
	\caption{The blue curve gives $E_w^{(3)}(A:B:C)$ and the red curves give $E_w(AB:C), E_w(AC:B), E_w(BC:A)$ respectively. In the notation of appendix \ref{app:signal} the blue triangle has edges $L_a,L_b,L_c$ and the red curves are $l_c,l_b,l_a$. One can see geometrically that $\Delta_w^{(3)}(A:B:C)$ defined in eq \ref{eq:delta3} is non-negative.}
	\label{fig:Delta3}
\end{figure}

In this section, we conjecture a new inequality for holographic states. Let us first introduce a modified version of the tripartite correlation signal $\Delta_w^{(3)}(A:B:C)$\cite{Bao:2025psl},
\begin{equation}\label{eq:delta3}
    \begin{aligned}
        &\Delta_w^{(3)}(A:B:C)\\
        & =E_w^{(3)}(A:B:C)\\
        & - \left(E_w(AB:C)+ E_w(AC:B)+ E_w(BC:A)\right),
    \end{aligned}
\end{equation}
where $E_w^{(3)}(A:B:C)$ is the tripartite entanglement-wedge cross section \cite{Umemoto:2018jpc,Bao:2018gck}, and $E_w(AB:C)$ is the usual bipartite entanglement wedge cross-section\cite{Nguyen:2017yqw,Takayanagi:2017knl} (see figure \ref{fig:Delta3} and appendix \ref{app:signal})

The proposed inequality for holographic states is
\begin{equation}\label{eq:new_conjecture}
    GM_R^{(3)}(A:B:C)\geq \frac{1}{2} \Delta_w^{(3)}(A:B:C),
\end{equation}
and since $\Delta_w^{(3)}(A:B:C)=0$ for pure states, it gives
\begin{equation}\label{eq:conjecture_valid_pure}
    GM^{(3)}(A:B:C)\geq 0,
\end{equation}
which is known to be true for holographic pure states\cite{Iizuka:2025caq}.

\begin{figure}[h!]
		\centering
		\includegraphics[width=0.40\textwidth]{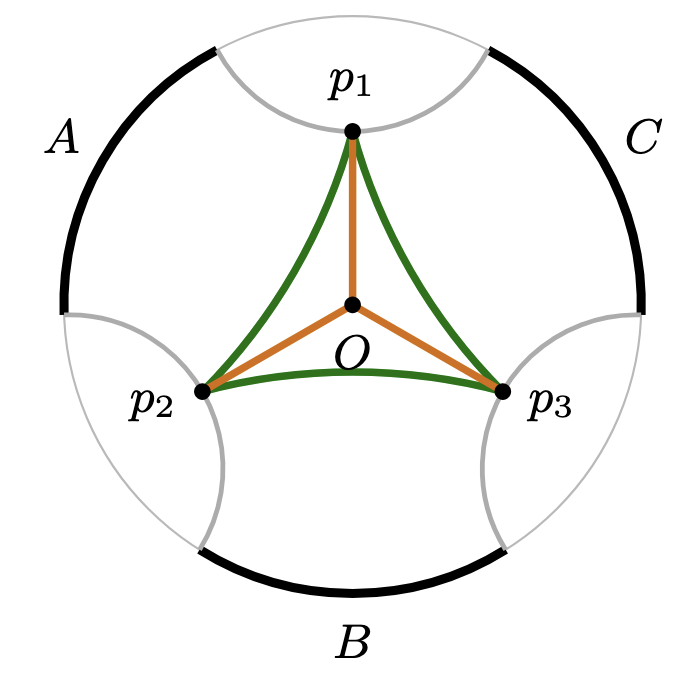}
	\caption{The orange tree gives $\frac{1}{2}S^{(3)}_R(A:B:C)$ and the green triangle gives $E_w^{(3)}(A:B:C)$. The three green curves meet pairwise at the vertices $p_1$, $p_2$ and $p_3$, at pairwise distances $\ell_{ij}$, and the orange tree joins the same three vertices through the trijunction $O$, its legs having lengths $|Op_i|$. The inequalities in eq \ref{eq:new_conjecture_sandwich} follow from the triangle inequality applied to these lengths.}
	\label{fig:geometric_proof}
\end{figure}

Let us consider a static time-slice in $AdS_3/CFT_2$, where we have the relation $2E_w(AB:C)=S^{(2)}_R(AB:C)$ \cite{Dutta:2019gen}. Writing $\sum_{\text{cyc}}E_w \equiv E_w(AB:C)+E_w(AC:B)+E_w(BC:A)$ and substituting this relation into the definition (\ref{eq:reflected-GM}) of the reflected genuine multi-entropy gives
\begin{equation}\label{eq:GM_R_in_Ew}
    GM^{(3)}_R(A:B:C)=\frac{1}{2}S^{(3)}_R(A:B:C)-\frac{1}{2}\sum_{\text{cyc}}E_w,
\end{equation}
while the definition (\ref{eq:delta3}) of the correlation signal gives
\begin{equation}\label{eq:Delta_in_Ew}
    \frac{1}{2}\Delta_w^{(3)}(A:B:C)=\frac{1}{2}E_w^{(3)}(A:B:C)-\frac{1}{2}\sum_{\text{cyc}}E_w.
\end{equation}
The bipartite cross-sections enter (\ref{eq:GM_R_in_Ew}) and (\ref{eq:Delta_in_Ew}) with the same coefficient and therefore cancel between the two sides of (\ref{eq:new_conjecture}), which is thus equivalent to
\begin{equation}\label{eq:new_conjecture_red}
    S^{(3)}_R(A:B:C) \geq E_w^{(3)}(A:B:C).
\end{equation}
In fact, for such geometric states (fig \ref{fig:geometric_proof}), the stronger sandwich inequality holds,
\begin{equation}\label{eq:new_conjecture_sandwich}
    2E_w^{(3)}(A:B:C) \geq S^{(3)}_R(A:B:C) \geq E_w^{(3)}(A:B:C),
\end{equation}
proved in appendix \ref{app:sandwich}, and reducing for pure states to
\begin{equation}\label{eq:new_conjecture_sandwich_pure}
    E_w^{(3)}(A:B:C) \geq S^{(3)}(A:B:C) \geq \frac{1}{2}E_w^{(3)}(A:B:C).
\end{equation}

In the geometry of figure \ref{fig:geometric_proof}, write $\ell_{ij}$ for the sides of the geodesic triangle and $|Op_i|$ for the legs of the tree, so that $E^{(3)}_w=\ell_{12}+\ell_{23}+\ell_{31}$ and $\frac{1}{2}S^{(3)}_R=|Op_1|+|Op_2|+|Op_3|$. Both bounds in (\ref{eq:new_conjecture_sandwich}) are then triangle inequalities. Since $Op_i\cup Op_j$ joins $p_i$ to $p_j$, every side obeys $\ell_{ij}\leq|Op_i|+|Op_j|$, and summing over the three pairs counts each leg twice, giving $E^{(3)}_w\leq S^{(3)}_R$. For the other bound, take the trijunction at the midpoint $m$ of $p_1p_2$. Each $p_i$ lies on $\gamma_i$, so the resulting tree is at most $d(m,p_1)+d(m,p_2)+d(m,p_3)=\ell_{12}+d(m,p_3)$, and convexity of the distance along a geodesic gives $d(m,p_3)\leq\frac{1}{2}(\ell_{13}+\ell_{23})$. Minimality then yields $\frac{1}{2}S^{(3)}_R\leq\ell_{12}+\frac{1}{2}(\ell_{13}+\ell_{23})\leq E^{(3)}_w$, i.e. $S^{(3)}_R\leq2E^{(3)}_w$. See Appendix \ref{app:sandwich} for more details.

\section{Discussion}\label{sec:discussion}
We have generalized the holographic signal inequality for mixed states by canonical purification in two different ways. In the minimal extension, we worked with the density matrix $\rho_{ABC}$ and computed the reflected genuine multi-entropy \cite{Yuan:2024yfg} and showed that this generalized inequality (\ref{eq:signal_ineq_mixed}) can be violated by some holographic states. The cause of violation in the minimal case is due to the fact that the Markov gap failed to capture tripartite entanglement from $\rho_{AB}$.

In the symmetric extension, we worked with the canonical purification of the density matrix $\rho_{ABC}$ and proposed a mixed-state generalization of the HSI that has a closer resemblance with the pure state HSI.

The authors in \cite{Balasubramanian:2025hxg} also suggested a four-party inequality using the genuine quadripartite multi-entropy $GM^{(4)}(A:B:C:D)\big|_{a=1/3}$. At four parties $GM^{(4)}$ is not unique: it carries a one-parameter ambiguity $a$, because the tripartite information is itself a signal of quadripartite entanglement and may be added with an arbitrary coefficient \cite{Iizuka:2025caq}. The inequality holds for the member $a=1/3$,
\begin{equation}\label{eq:four-party-signal-ineq}
\begin{aligned}
    & \frac{1}{2}\left(R^{(3)}(A:B) + R^{(3)}(C:D)\right)\\
    & \geq GM^{(4)}(A:B:C:D)\big|_{a=1/3}+\frac{1}{3}\sum GM^{(3)}\\
    & =S^{(4)}(A:B:C:D)- \frac{S(A)+S(B)+S(C)+S(D)}{2},
\end{aligned}
\end{equation}
where the sum in the second line runs over all distinct tripartitions of $ABCD$, i.e. over the six terms $GM^{(3)}(AB:C:D)$, $GM^{(3)}(AC:B:D)$, $GM^{(3)}(AD:B:C)$, $GM^{(3)}(BC:A:D)$, $GM^{(3)}(BD:A:C)$ and $GM^{(3)}(CD:A:B)$, and we have used the simplifications for pure states to get the equality in the third line.

The same canonical-purification argument applied at four parties leads to an analogous mixed-state generalization $GM^{(4)}_R(A:B:C:D)$. One can choose a boundary configuration such that entanglement wedges of $AB$ and $CD$ are disconnected, but the entanglement wedge of $ABCD$ is connected. If using the minimal extension, the LHS is zero again, whereas the RHS can be positive.

Unlike the bipartite case, multipartite entanglement has a rich SLOCC classification. While three-qubit states fall into either the GHZ or W class, there are twelve types of three-party qutrit entanglement \cite{Yang:2008vwi} and nine types of four-party qubit entanglement \cite{Verstraete:2002gqj}. It is interesting to check how many of them satisfy a Renyi version of the holographic signal inequality. Unlike the GHZ state, computing the multi-entropy at $n\rightarrow 1$ may not be straightforward for many (if not all) of these states. In such a case, one also needs to determine (and prove) the monotonicity of Renyi multi-entropy in the Renyi parameter $n$, which is known to hold for ordinary Renyi entropies. We will address these questions in a future work \cite{DN:preparation}.

The proposed new inequality eq (\ref{eq:new_conjecture}) and its simplified version eq (\ref{eq:new_conjecture_red}) are constructed using holographic quantities. One may ask whether we can write an analogue of this inequality for more general quantum systems. It is possible to replace the entanglement wedge cross-section with the entanglement of purification, which gives us
\begin{equation}
\label{eq:new_conjecture_purification}
    GM_R^{(3)}(A:B:C)\geq \frac{1}{2}\Delta^{(3)}_p(A:B:C),
\end{equation}
where
\begin{equation}
    \begin{aligned}
        &\Delta_p^{(3)}(A:B:C)\\
        & =E_p^{(3)}(A:B:C)\\
        & - \left(E_p(AB:C)+ E_p(AC:B)+ E_p(BC:A)\right).
    \end{aligned}
\end{equation}
For pure tripartite states, $\Delta_p^{(3)}=0$\cite{Bao:2025psl}, so this inequality simplifies to 
\begin{equation}\label{eq:new_conj_pur_red}
GM^{(3)}(A:B:C)\geq 0.    
\end{equation}

Note that (\ref{eq:new_conj_pur_red}) is not known to hold for general quantum states --- as remarked in section \ref{sec:mixed}, $GM^{(3)}\geq0$ has been established in holography but not in general. The conjecture (\ref{eq:new_conjecture_purification}) is therefore at least as strong as that open positivity statement, which is both what makes it interesting and the reason we state it only as a conjecture. It is interesting to understand the class of density matrices that satisfy it.

We have used the vacuum $AdS_3$ subregions for our mixed state. We expect similar arguments to hold for subregions in a black hole background. The violation of (\ref{eq:signal_ineq_mixed}) turns only on $M_{ABC}$ being connected while $M_{AB}$ is not, and a horizon supplies a further channel through which $M_{ABC}$ can connect without connecting $M_{AB}$.

\section*{Acknowledgements}
J.N. would like to thank Ning Bao and Norihiro Iizuka for comments. This research was supported in part by grant NSF PHY-2309135 to the Kavli Institute for Theoretical Physics (KITP) and the Graduate Assistantship at Northeastern University. Currently, the author is supported by Beijing Institute of Mathematical Sciences and Applications. No generative AI was used in writing the main text of the paper except Gemini and Claude to generate the figures. Claude was used to review some background materials in the appendix and the author takes responsibility for the content.

\appendix

\section{Multi-entropy}\label{app:multi-entropy}

We collect here the definitions used in section \ref{sec:review}, following \cite{Harper:2025uui}.

\emph{Multi-invariants.} Let $\ket{\psi}$ be the $\mathtt{q}$-partite pure state (\ref{eq:state-psi}). A local-unitary invariant is built from $s$ copies of $\psi$ and $s$ copies of $\bar\psi$ by contracting every index, the pattern of contraction being specified by one permutation $g_\mathtt{k}$ per party \cite{Gadde:2024taa},
\begin{equation}\label{eq:app-invariant}
\begin{aligned}
    \mathcal{E}(g_1,\dots,g_\mathtt{q})&=\Big(\prod_{r=1}^{s}\psi_{i^{(r)}_1\dots i^{(r)}_\mathtt{q}}\Big)\Big(\prod_{r=1}^{s}\bar{\psi}^{\,j^{(r)}_1\dots j^{(r)}_\mathtt{q}}\Big)\prod_{\mathtt{k}=1}^{\mathtt{q}}\delta^{\vec{i}_\mathtt{k}}_{g_\mathtt{k}\cdot\vec{j}_\mathtt{k}},\\
    \delta^{\vec{i}_\mathtt{k}}_{g_\mathtt{k}\cdot\vec{j}_\mathtt{k}}&\equiv\delta^{i^{(1)}_\mathtt{k}}_{j^{(g_\mathtt{k}\cdot1)}_\mathtt{k}}\cdots\,\delta^{i^{(s)}_\mathtt{k}}_{j^{(g_\mathtt{k}\cdot s)}_\mathtt{k}} .
\end{aligned}
\end{equation}
The $g_\mathtt{k}$ are taken from a group $G$ of order $s$ acting in its regular representation. Relabelling the $\psi$'s once the $\bar\psi$'s are fixed leaves $\mathcal{E}$ unchanged,
\begin{equation}\label{eq:app-redundancy}
    \mathcal{E}(g_1,\dots,g_\mathtt{q})=\mathcal{E}(g_1\,g,\dots,g_\mathtt{q}\,g),\qquad g\in S_s,
\end{equation}
so one element may always be set to the identity; this is the same as tracing out that party and working with the reduced density matrix of the remaining $\mathtt{q}-1$.

In a $2$d CFT the permutations are gluing instructions: cutting $s$ copies along the intervals at $t_E=0$ and attaching copy $r$ to copy $g_\mathtt{k}\cdot r$ across the interval assigned to party $\mathtt{k}$ produces a branched cover, and $\mathcal{E}$ is its partition function. Encircling the point where two intervals $\mathtt{k}$ and $\mathtt{k}'$ meet permutes the sheets by $g_\mathtt{k}g_{\mathtt{k}'}^{-1}$, so if that element has order $l$ the uniformising map takes the endpoint to a corner of angle $\pi/l$; the genus of the resulting surface is then fixed by Riemann-Hurwitz. Equivalently, $\mathcal{E}$ is a correlator of twist operators $\sigma_{h}$, one at each interval endpoint, labelled by the corresponding monodromy.

\emph{Multi-entropy.} The multi-entropy is the choice $G=\mathbb{Z}_n^{\mathtt{q}-1}$, for which every twist operator has the same cycle length $n$ \cite{Gadde:2022cqi,Penington:2022dhr}. For $\mathtt{q}=3$ we take $\mathbb{Z}_n^2=\langle a,b\,|\,a^n=b^n=e,\ ab=ba\rangle$ acting on $s=n^2$ copies by
\begin{equation}\label{eq:app-ab}
\begin{aligned}
    a&:(1,\dots,n)(n+1,\dots,2n)\cdots(n^2-n+1,\dots,n^2),\\
    b&:(1,n+1,\dots,n^2-n+1)\cdots(n,2n,\dots,n^2),
\end{aligned}
\end{equation}
and assign
\begin{equation}\label{eq:app-assign}
    g_C=e,\qquad g_A=a,\qquad g_B=b,
\end{equation}
the trivial $g_C$ being the gauge choice (\ref{eq:app-redundancy}). The R\'enyi multi-entropy is this invariant with a fixed normalisation,
\begin{equation}\label{eq:app-ME}
    S^{(3)}_n(A:B:C)=\frac{1}{1-n}\frac{1}{n}\log\mathcal{E}\big(e,a,b\big),
\end{equation}
which is (\ref{eq:renyi-3-multi}); for general $\mathtt{q}$ the prefactor is $\frac{1}{1-n}\frac{1}{n^{\mathtt{q}-2}}$. At $n=2$ the group elements (\ref{eq:app-ab}) read $a=(1\,2)(3\,4)$ and $b=(1\,3)(2\,4)$ on four copies, and (\ref{eq:app-invariant}) reduces to the contraction $\varepsilon$ quoted in section \ref{sec:review}.

For three adjacent intervals $A=[x_1,x_2]$, $B=[x_2,x_3]$ in the vacuum of a $2$d CFT, conformal symmetry fixes
\begin{equation}\label{eq:app-Sn3-cft}
    S^{(3)}_n=\frac{c}{12}\Big(1+\frac{1}{n}\Big)\log\frac{x_{21}x_{31}x_{32}}{\epsilon^{3}}+\kappa^{(3)}_n ,
\end{equation}
with $x_{ij}=x_i-x_j$ and the cutoff $\epsilon$ fixed by the corresponding convention for the R\'enyi entropy. The constant is $\kappa^{(3)}_2=\frac{c}{4}\log2$, and the holographic formula predicts $\kappa^{(3)}_1=\frac{c}{2}\log\frac{2}{\sqrt3}$ \cite{Harper:2025uui}.

\section{Genuine multi-entropy}\label{app:genuine}

The multi-entropy $S^{(\mathtt{q})}_n$ is invariant under local unitaries, symmetric in the parties, and additive on tensor products, but it is not by itself a measure of genuine $\mathtt{q}$-partite correlation: it is also sensitive to $\tilde{\mathtt{q}}$-partite entanglement for every $\tilde{\mathtt{q}}<\mathtt{q}$. Following \cite{Iizuka:2025caq,Iizuka:2025ioc}, the genuine multi-entropy $GM^{(\mathtt{q})}_n$ is defined as the symmetric linear combination of multi-entropies that (i) contains $S^{(\mathtt{q})}_n$ and (ii) vanishes on every factorised state
\begin{equation}\label{eq:app-factorised}
    \ket{\psi}_{A_1\cdots A_\mathtt{q}}=\ket{\psi}_{A_1\cdots A_{\tilde{\mathtt{q}}}}\otimes\ket{\psi}_{A_{\tilde{\mathtt{q}}+1}\cdots A_\mathtt{q}},\qquad \tilde{\mathtt{q}}=1,\dots,\mathtt{q}-1,
\end{equation}
condition (ii) being consistent precisely because $S^{(\mathtt{q})}_n$ is additive. For $\mathtt{q}=3$ the combination is unique and is the one used in (\ref{eq:GME-def1}); since $S^{(2)}_n(AB:C)=S^{(2)}_n(C)$ for a pure state, it may equivalently be written as
\begin{equation}\label{eq:app-GM3-alt}
    GM^{(3)}_n(A:B:C)=S^{(3)}_n(A:B:C)-\frac{1}{2}\sum_{X=A,B,C}S^{(2)}_n(X),
\end{equation}
which makes the removal of all bipartite contributions manifest.

Holographically, $S^{(\mathtt{q})}$ is the area of the minimal $\mathtt{q}$-way cut, and the subtraction in (\ref{eq:app-GM3-alt}) cancels the cutoff dependence of that cut against the Ryu-Takayanagi surfaces, leaving a finite quantity. In $AdS_3$ one has the sharper statement \cite{Iizuka:2025caq}
\begin{equation}\label{eq:app-bound}
\begin{aligned}
    S^{(\mathtt{q})}-\frac{1}{2}\sum_{\mathtt{k}}S^{(2)}(A_\mathtt{k})\;\geq\;\frac{3}{4G_N}\log\frac{2}{\sqrt3}\,N_V ,
\end{aligned}
\end{equation}
where $N_V$ is the number of vertices of the minimal $\mathtt{q}$-way cut. For $\mathtt{q}=3$ that cut is the triway cut of (\ref{eq:multi-holo}), which has a single vertex --- the trijunction $O$ --- so (\ref{eq:app-bound}) is exactly the bound (\ref{eq:GME_AdS3}) quoted in the main text. The bound is saturated by three adjacent intervals in the vacuum, where (\ref{eq:app-Sn3-cft}) gives $GM^{(3)}=\kappa^{(3)}_1=\frac{c}{2}\log\frac{2}{\sqrt3}=\frac{3}{4G_N}\log\frac{2}{\sqrt3}$ by Brown-Henneaux. When the subregions are far apart the cut degenerates, the vertex count drops to zero, and the bound becomes empty --- consistent with $GM^{(3)}_R=0$ in the disconnected phase of section \ref{sec:mixed}.

\section{The tripartite correlation signal}\label{app:signal}

We summarise here the construction of \cite{Bao:2025psl} used in section \ref{sec:conj}.

Following \cite{Umemoto:2018jpc}, the multipartite entanglement wedge cross-section of $n$ boundary subregions is obtained by partitioning the boundary of their entanglement wedge,
\begin{equation}\label{eq:app-partition}
    \partial M_{A_1\cdots A_n}=\tilde{A}_1\cup\cdots\cup\tilde{A}_n,\qquad \tilde{A}_m\supseteq A_m,
\end{equation}
setting $\Sigma^{(n)}_{\min}=\bigcup_m\Sigma_{\min}(\tilde{A}_m)$ with $\Sigma_{\min}(\tilde{A}_m)$ minimal and homologous to $\tilde{A}_m$, and minimising its area over the choice of partition,
\begin{equation}\label{eq:app-MEWCS}
    E^{(n)}_w(A_1:\cdots:A_n)=\min_{\{\tilde{A}_m\}}\frac{L_{\rm AdS}}{4G_N}\,\mathrm{Area}\big(\Sigma^{(n)}_{\min}\big).
\end{equation}
With the convention $E^{(2)}_w(A:B):=2E_w(A:B)$, the tripartite correlation signal is
\begin{equation}\label{eq:app-delta}
    \Delta^{(3)}_w(A:B:C)=E^{(3)}_w(A:B:C)-\frac{1}{2}\sum_{\text{cyc}}E^{(2)}_w ,
\end{equation}
which is (\ref{eq:delta3}). By the conjecture $E^{(n)}_p=E^{(n)}_w$ it coincides with the correlation signal $\Delta^{(3)}_p$ built from the multipartite entanglement of purification.

On a static slice of $AdS_3$ with $M_{ABC}$ connected, both objects in (\ref{eq:app-delta}) are drawn in figure \ref{fig:Delta3}: write $l_a$ for the red curve separating $A$ from $BC$, and $L_a,L_b,L_c$ for the three arcs of the blue tripartite cross-section, so that
\begin{equation}\label{eq:app-lL}
    E^{(2)}_w(A:BC)=2l_a,\qquad E^{(3)}_w(A:B:C)=L_a+L_b+L_c,
\end{equation}
and $\Delta^{(3)}_w=\sum_i L_i-\sum_i l_i$. The $L_i$ close into a geodesic triangle with one vertex on each component of $\Gamma^{(3)}_{\min}(A:B:C)$, whereas each $l_i$ is the unique common perpendicular between the corresponding pair of components. Since a hyperbolic triangle with vertices in the interior has strictly positive angles, the $l_i$ cannot themselves form a triangle, and at least two of the $L_i$ strictly exceed the corresponding $l_i$; hence $\Delta^{(3)}_w\geq0$ \cite{Bao:2025psl}. It vanishes in two limits: when $M_{ABC}$ is disconnected, where all six quantities degenerate, and when $\rho_{ABC}$ is pure, where $L_i=l_i=\Gamma_{\min}$ for each party --- the statement used below (\ref{eq:new_conjecture}).

\section{The $C$-excised construction}\label{app:c-excised}

The symmetric extension of section \ref{sec:mixed} involves two canonical purifications in succession, and the quantities $S_{RR}(A:B;[C])$ and $I_R(A:B;[C])$ entering (\ref{eq:reflected-residual-entropy}) were quoted there without construction. We supply it here.

\emph{The doubled geometry.} The entanglement wedge $M_{ABC}$ of $\rho_{ABC}$ is bounded by $A\cup B\cup C$ together with the Ryu-Takayanagi surface $\gamma_{ABC}$, which in the connected phase has three components. The canonical purification (\ref{eq:rho_can_pur}) glues a second copy of $M_{ABC}$ to the first along $\gamma_{ABC}$. The conformal boundary of the result carries the three doubled regions $AA^{*}$, $BB^{*}$ and $CC^{*}$, one for each party, and $\ket{\sqrt{\rho_{ABC}}}$ is pure on their union.

\begin{figure*}[t]
    \centering
    \begin{subfigure}[b]{0.31\textwidth}\centering
      \includegraphics[width=\textwidth]{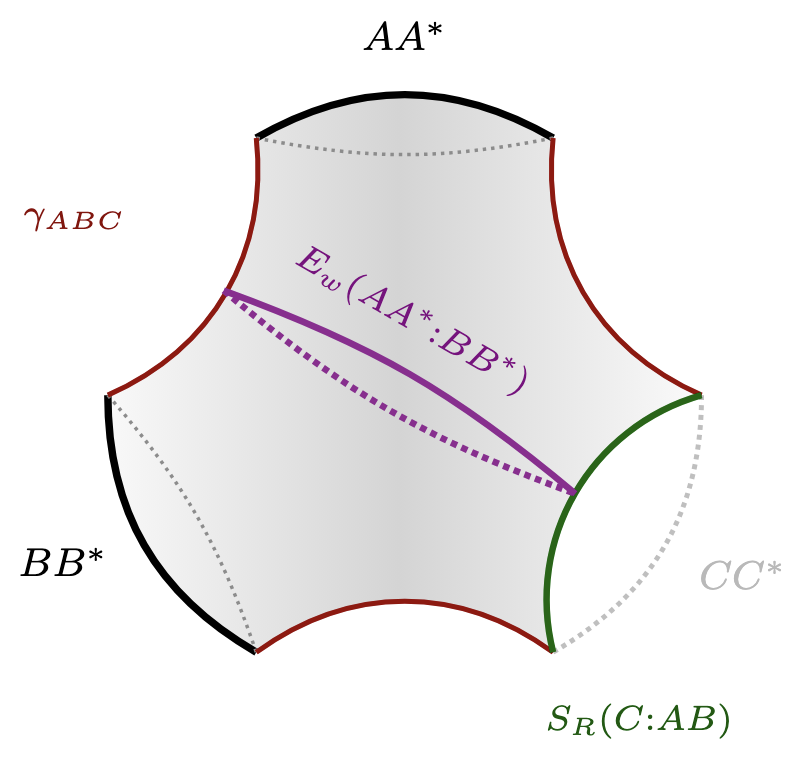}
      \caption{}\label{fig:cexc-wedge}
    \end{subfigure}\hspace{0.04\textwidth}
    \begin{subfigure}[b]{0.42\textwidth}\centering
      \includegraphics[width=\textwidth]{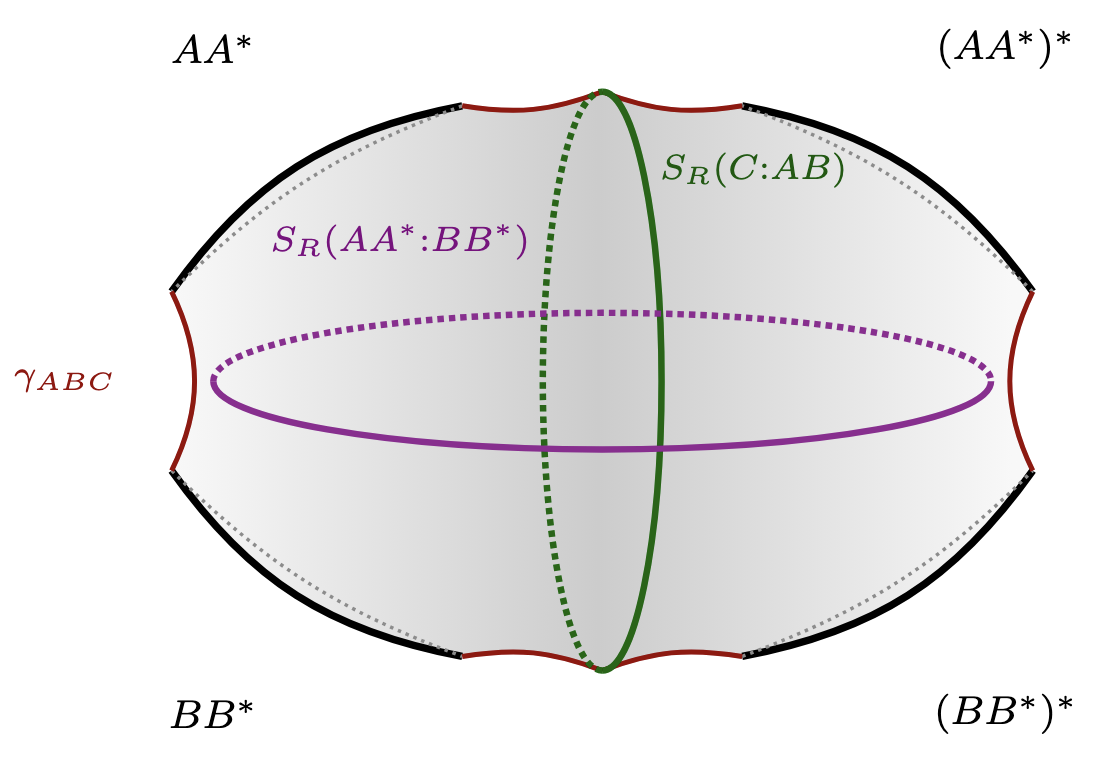}
      \caption{}\label{fig:cexc-worm}
    \end{subfigure}
  \caption{(a) The $C$-excised wedge. The canonical purification $\ket{\sqrt{\rho_{ABC}}}$ glues two copies of $M_{ABC}$ along the three components of $\gamma_{ABC}$ (red), so that its boundary carries $AA^{*}$, $BB^{*}$ and $CC^{*}$; tracing out $CC^{*}$ removes the dotted region beyond the green surface, which is the Ryu-Takayanagi surface of $AA^{*}BB^{*}$ and equally that of $CC^{*}$. The violet curve, running from $\gamma_{ABC}$ to the green surface, is the cross-section of what remains, and it joins $AA^{*}$ to $BB^{*}$ even when the entanglement wedge of $AB$ is disconnected. (b) The second canonical purification, gluing two copies of (a) along that green surface, drawn here as a wormhole with the throat shaded and the far side of each surface dotted. The violet closed curve is the reflected entropy surface $S_R(AA^{*}{:}BB^{*})=2E_w(AA^{*}{:}BB^{*})$; projected onto a single copy of $M_{ABC}$ it becomes the curve $\Gamma(A{:}B|C)$ of figure \ref{eq:C-excised-reflected}.}
    \label{fig:app-cexcised}
\end{figure*}

\emph{Excising $C$.} Tracing out $CC^{*}$ leaves $\hat{\rho}_{AB}^{A^{*}B^{*}}$, whose entanglement wedge is bounded by $AA^{*}$, $BB^{*}$ and the Ryu-Takayanagi surface of $AA^{*}BB^{*}$. Since $\ket{\sqrt{\rho_{ABC}}}$ is pure, that surface is the one homologous to $CC^{*}$, namely the reflected entropy surface of $C$ against $AB$. This is the \emph{$C$-excised} wedge (figure \ref{fig:cexc-wedge}). The essential point is that $AA^{*}$ and $BB^{*}$ are joined through it even when the entanglement wedge of $AB$ in the original geometry is disconnected: the doubling connects them through the region vacated by $CC^{*}$.

\emph{The $C$-excised mutual information.} Each term of $I_R(A:B;[C])=I(AA^{*}:BB^{*})$ in (\ref{eq:sym_MI}) is an ordinary entanglement entropy in $\ket{\sqrt{\rho_{ABC}}}$, and by construction $S(AA^{*})=S^{(2)}_R(A:BC)=2E_w(A:BC)$, with the same statement for $BB^{*}$. Purity lets us trade the third term for its complement, $S(AA^{*}BB^{*})=S(CC^{*})=2E_w(C:AB)$, so
\begin{equation}\label{eq:app-J}
\begin{aligned}
    \mathcal{J}\;\equiv\;&E_w(A{:}BC)+E_w(B{:}AC)\\&-E_w(C{:}AB),
\end{aligned}
\end{equation}
\begin{equation}\label{eq:app-IR}
    I_R(A:B;[C])=2\,\mathcal{J},
\end{equation}
manifestly finite, and non-zero whenever $M_{ABC}$ is connected --- in particular for the geometry of figure \ref{fig:example_geometry_violate}, where $I(A:B)$ itself vanishes.

\emph{The $C$-excised reflected entropy.} Purifying $\hat{\rho}_{AB}^{A^{*}B^{*}}$ in turn (figure \ref{fig:cexc-worm}) and using $S_R=2E_w$ gives $S_{RR}(A:B;[C])=S_R(AA^{*}:BB^{*})=2E_w(AA^{*}:BB^{*})$. Projecting onto a single copy of $M_{ABC}$ with the $\mathbb{Z}_2$ symmetry of the first purification, this cross-section becomes the curve $\Gamma(A:B|C)$ of figure \ref{eq:C-excised-reflected}, running from the component of $\gamma_{ABC}$ that separates $A$ from $B$ to the bipartite cross-section $E_w(C:AB)$. The doubled cross-section consists of two mirror copies of it, so $E_w(AA^{*}:BB^{*})=2\Gamma(A:B|C)$ and
\begin{equation}\label{eq:app-SRR}
    S_{RR}(A:B;[C])=4\,\Gamma(A:B|C).
\end{equation}

\emph{Result.} Combining (\ref{eq:app-IR}) and (\ref{eq:app-SRR}) with the definition (\ref{eq:reflected-residual-entropy}),
\begin{equation}\label{eq:app-RR}
    R^{(3)}_R(A:B;[C])=2\,\Gamma(A:B|C)-\mathcal{J},
\end{equation}
a purely geometric expression in which $\Gamma(A:B|C)$ is the only new object. As a check, in the limit where the gaps close and $\rho_{ABC}$ becomes pure the doubling is trivial: $\Gamma(A:B|C)\rightarrow\Gamma_{AB}$ and $E_w(X{:}YZ)\rightarrow S(X)$, so $\mathcal{J}\rightarrow S(A)+S(B)-S(C)$ and (\ref{eq:app-RR}) reduces to $2\Gamma_{AB}-\big(S(A)+S(B)-S(C)\big)$, which is the Markov gap of (\ref{eq:Markov_gap}) and confirms that the symmetric extension (\ref{eq:sym-ext-HSI-single}) reduces to the pure-state inequality (\ref{eq:signal_ineq_pure}).

\section{HSI violation in vacuum $AdS_3$}\label{app:explicit}

We evaluate both sides of the minimally extended inequality (\ref{eq:signal_ineq_mixed}) for the symmetric configuration of \cite{Bao:2025psl}: three equal intervals of half-width $\alpha$ with centres $2\pi/3$ apart on the boundary of vacuum $AdS_3$, so that $\alpha\in(0,\pi/3)$ and $\alpha\rightarrow\pi/3$ is the pure-state limit. We work throughout on the Poincar\'e disk with $L_{\rm AdS}=1$ and quote all quantities in units $L_{\rm AdS}/4G_N=1$.

\emph{One building block.} Every surface below is fixed by a single function. The geodesic subtending a boundary arc of half-width $h$ lies at hyperbolic distance
\begin{equation}\label{eq:app-rho}
    \rho(h)=\log\cot\frac{h}{2}
\end{equation}
from the centre of the disk. The three components of $\gamma_{ABC}$ subtend the gaps, of half-width $\pi/3-\alpha$, while the far geodesic bounding the connected wedge of $AB$ subtends an arc of half-width $2\pi/3-\alpha$; we abbreviate
\begin{equation}\label{eq:app-rho12}
    \rho_1=\rho\!\left(\tfrac{\pi}{3}-\alpha\right),\qquad
    \rho_2=\rho\!\left(\tfrac{2\pi}{3}-\alpha\right).
\end{equation}
Comparing the connected and disconnected candidate surfaces, the wedge $M_{ABC}$ is connected for $\alpha>\pi/6$.

\emph{Right-hand side.} By the $\mathbb{Z}_2$ symmetry of the canonical purification the triway cut computing $S^{(3)}_R$ consists of two mirror copies of a tree in a single copy of $M_{ABC}$, each leg meeting its component of $\gamma_{ABC}$ perpendicularly. Symmetry places the trijunction at the centre, so each leg has length $\rho_1$ and
\begin{equation}\label{eq:app-SR3}
    \tfrac{1}{2}S^{(3)}_R(A:B:C)=3\rho_1 .
\end{equation}
The bipartite cross-section $E_w(A:BC)$ is the common perpendicular of the two components of $\gamma_{ABC}$ bounding $A$, whose normals subtend an angle $2\pi/3$; contracting the corresponding unit normals gives
\begin{equation}\label{eq:app-EwABC}
    E_w(A:BC)=\operatorname{arccosh}\!\left(\tfrac{3}{2}\cosh^{2}\!\rho_1-1\right),
\end{equation}
and likewise for the two permutations. Using $S^{(2)}_R=2E_w$ in (\ref{eq:reflected-GM}), the right-hand side of (\ref{eq:signal_ineq_mixed}) is
\begin{equation}\label{eq:app-RHS}
    GM^{(3)}_R=3\rho_1-\tfrac{3}{2}\operatorname{arccosh}\!\left(\tfrac{3}{2}\cosh^{2}\!\rho_1-1\right)
\end{equation}
for $\alpha>\pi/6$, and zero below. As $\alpha\rightarrow\pi/3$ one has $\rho_1\rightarrow\infty$ and $\operatorname{arccosh}(\tfrac32\cosh^2\rho_1-1)\rightarrow 2\rho_1+\log\tfrac{3}{4}$, so that $GM^{(3)}_R\rightarrow\tfrac{3}{2}\log\tfrac{4}{3}=3\log\tfrac{2}{\sqrt3}$, saturating the bound quoted in section \ref{sec:mixed}.

\emph{Left-hand side.} The connected wedge of $AB$ is bounded by the two geodesics at distances $\rho_1$ and $\rho_2$ on opposite sides of the centre, so their common perpendicular is the diameter joining them and
\begin{equation}\label{eq:app-EwAB}
    E_w(A:B)=\rho_1+\rho_2 .
\end{equation}
Writing $u=\tfrac{1}{2}(\rho_1+\rho_2)$, the mutual information in the same phase is $I(A:B)=4\log\sinh u$, so that with $S_R=2E_w$
\begin{equation}\label{eq:app-LHS}
    \tfrac{1}{2}R^{(3)}(A:B)=2u-2\log\sinh u .
\end{equation}
The expression is valid only in the connected phase, which requires $I(A:B)>0$, i.e. $\sinh u>1$; the wedge of $AB$ therefore connects at
\begin{equation}\label{eq:app-alphastar}
\begin{aligned}
    \rho_1+\rho_2&=2\operatorname{arcsinh}1=2\log\!\left(1+\sqrt2\right),\\
    \text{i.e.}\qquad \alpha_{*}&=\arcsin\sqrt{\tfrac{3}{8}}\simeq0.659 ,
\end{aligned}
\end{equation}
where the left-hand side jumps from zero to $2\log(1+\sqrt2)$ and thereafter decreases monotonically to $2\log2$ as $\alpha\rightarrow\pi/3^{-}$.

\begin{figure}[h!]
    \centering
    \includegraphics[width=0.46\textwidth]{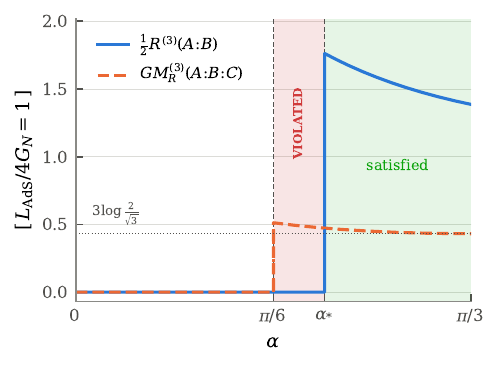}
  \caption{The two sides of (\ref{eq:signal_ineq_mixed}) for three equal intervals of half-width $\alpha$ in vacuum $AdS_3$, in units $L_{\rm AdS}/4G_N=1$. Below $\alpha=\pi/6$ neither wedge is connected and both sides vanish. Between $\pi/6$ and $\alpha_{*}=\arcsin\sqrt{3/8}$ the wedge of $ABC$ has connected while that of $AB$ has not, so $GM^{(3)}_R>0$ while the Markov gap is still identically zero and the inequality is violated. At $\alpha_{*}$ the Markov gap turns on discontinuously at $2\log(1+\sqrt2)$ and the inequality is satisfied thereafter, both sides approaching their limiting values $2\log2$ and $3\log\frac{2}{\sqrt3}$ as $\alpha\rightarrow\pi/3^{-}$.}
    \label{fig:lhs_rhs}
\end{figure}

\emph{The violation.} Both sides are plotted in figure \ref{fig:lhs_rhs}. For $\alpha<\pi/6$ they vanish together. At $\alpha=\pi/6$ the right-hand side jumps to
\begin{equation}
    GM^{(3)}_R\big|_{\pi/6^{+}}=3\log\frac{2+\sqrt3}{\sqrt2+\sqrt3}\simeq0.512 ,
\end{equation}
while the left-hand side is still identically zero, and it remains zero until $\alpha_{*}$. The inequality is therefore violated throughout
\begin{equation}\label{eq:app-window}
    \frac{\pi}{6}<\alpha<\alpha_{*},
\end{equation}
by a margin that decreases from $0.512$ to $0.473$ across the window. Above $\alpha_{*}$ the left-hand side exceeds the right by at least $2\log2-3\log\tfrac{2}{\sqrt3}\simeq0.955$, so the inequality is comfortably satisfied throughout $\alpha_{*}<\alpha<\pi/3$. The endpoint itself is not attained: for every $\alpha<\pi/3$ the region $A\cup B$ is disconnected and $\Gamma_{AB}$ has both ends on $\gamma_{ABC}$, whereas at $\alpha=\pi/3$ it is a single interval, $\Gamma_{AB}$ loses one of those two anchors, and $\tfrac{1}{2}R^{(3)}(A:B)$ drops discontinuously from its limiting value $2\log2$ to the pure-state value $\log2$ of \cite{Balasubramanian:2025hxg}.

\section{Proof of the sandwich inequality}\label{app:sandwich}

The argument given below (\ref{eq:new_conjecture_sandwich_pure}) in the main text assumes that the vertices of the geodesic triangle coincide with the endpoints of the tree. That holds when $\rho_{ABC}$ is pure, since both are then anchored on the same boundary points, and it survives in sufficiently symmetric mixed configurations such as the one of appendix \ref{app:explicit}; it fails in general. For a generic mixed state the two objects are anchored on the components $\gamma_1,\gamma_2,\gamma_3$ of $\Gamma^{(3)}_{\min}(A:B:C)$ and are obtained from \emph{independent} minimisations,
\begin{equation}\label{eq:app-PT}
\begin{aligned}
    E^{(3)}_w&=P:=\min_{p_i\in\gamma_i}\ \sum_{i<j}d(p_i,p_j),\\
    \tfrac{1}{2}S^{(3)}_R&=T:=\min_{O\in M_{ABC}}\ \sum_{i=1}^{3}d(O,\gamma_i),
\end{aligned}
\end{equation}
the first over the three vertices, the second over the trijunction. Both minima are attained, and only the triangle inequality for the induced distance $d$ on the time slice is needed below; in particular the argument does not use vacuum $AdS_3$.

\emph{Lower bound.} Let $O^{\ast}$ attain $T$ and let $q_i\in\gamma_i$ be the foot of the perpendicular dropped from $O^{\ast}$, so that $T=\sum_i d(O^{\ast},q_i)$. The triple $\{q_i\}$ is admissible in the first line of (\ref{eq:app-PT}), whence
\begin{equation}\label{eq:app-lem1}
    P\ \leq\ \sum_{i<j}d(q_i,q_j)\ \leq\ \sum_{i<j}\big[d(q_i,O^{\ast})+d(O^{\ast},q_j)\big]\ =\ 2T,
\end{equation}
each leg being counted exactly twice in the double sum. This is $E^{(3)}_w\leq S^{(3)}_R$, the second inequality of (\ref{eq:new_conjecture_sandwich}); by the reduction of (\ref{eq:new_conjecture}) to (\ref{eq:new_conjecture_red}) it establishes (\ref{eq:new_conjecture}) for every configuration to which (\ref{eq:app-PT}) applies. We stress what that parametrisation assumes, since it is where the conjecture reaches beyond the argument: that the entanglement wedge $M_{ABC}$ is connected with $\gamma_{ABC}$ having exactly three components; that the minimal web computing $S^{(3)}_R$ has a single trijunction in the single copy, as proposed in \cite{Yuan:2024yfg}; that the minimal tripartite cross-section is a triangle with one vertex on each component, as proposed in \cite{Umemoto:2018jpc}; and that both proposed holographic duals hold. Geometries in which the minimal web or the minimal cross-section has a different combinatorial structure --- more components of $\gamma_{ABC}$, additional junctions, or higher bulk dimension --- are not covered, and we do not have an argument there. This is why (\ref{eq:new_conjecture}) is offered as a conjecture rather than a theorem.

\emph{Upper bound.} Let $\{p_i\}$ attain $P$, write $\ell_{ij}=d(p_i,p_j)$, and let $m_{ij}$ be the midpoint of the geodesic segment $p_ip_j$. Each choice $O=m_{ij}$ is admissible in the second line of (\ref{eq:app-PT}), and since $p_i\in\gamma_i$,
\begin{equation}\label{eq:app-mid}
    T\ \leq\ \sum_{i}d(m_{12},p_i)\ =\ \ell_{12}+d(m_{12},p_3)\ \leq\ \ell_{12}+\tfrac{1}{2}\big(\ell_{13}+\ell_{23}\big),
\end{equation}
the last step by convexity of $d(\cdot,p_3)$ along $p_1p_2$. Averaging the three such choices,
\begin{equation}\label{eq:app-lem2}
    3T\ \leq\ \sum_{i<j}\Big[\ell_{ij}+\tfrac{1}{2}\big(\ell_{ik}+\ell_{jk}\big)\Big]\ =\ 2P .
\end{equation}
Therefore $S^{(3)}_R\leq\tfrac{4}{3}E^{(3)}_w$, which is stronger than the first inequality of (\ref{eq:new_conjecture_sandwich}); retaining only the single choice (\ref{eq:app-mid}) returns the weaker form, since $\tfrac12(\ell_{13}+\ell_{23})\leq\ell_{13}+\ell_{23}$. When $M_{ABC}$ is disconnected both sides vanish and the statement is empty.

\emph{Explicit check.} In the symmetric configuration of appendix \ref{app:explicit} both quantities are functions of the single distance $\rho_1$ of (\ref{eq:app-rho12}). The tree has three legs of length $\rho_1$, while the minimal triangle has its vertices at the points of $\gamma_1,\gamma_2,\gamma_3$ closest to the centre, so that
\begin{equation}\label{eq:app-PTexplicit}
    T=3\rho_1,\qquad
    P=3\operatorname{arccosh}\!\left(\tfrac{3}{2}\cosh^{2}\!\rho_1-\tfrac{1}{2}\right).
\end{equation}
Comparing with (\ref{eq:app-EwABC}) gives the correlation signal of appendix \ref{app:signal} in closed form,
\begin{equation}\label{eq:app-Delta}
\begin{aligned}
    \Delta^{(3)}_w=3\Big[&\operatorname{arccosh}\!\left(\tfrac{3}{2}\cosh^{2}\!\rho_1-\tfrac{1}{2}\right)\\
    &-\operatorname{arccosh}\!\left(\tfrac{3}{2}\cosh^{2}\!\rho_1-1\right)\Big]>0,
\end{aligned}
\end{equation}
manifestly positive because $\operatorname{arccosh}$ is increasing. Both (\ref{eq:app-lem1}) and (\ref{eq:app-lem2}) hold strictly throughout $\pi/6<\alpha<\pi/3$. The second comes closest to saturation at the lower end, where $3T/2P=3\log(2+\sqrt3)\big/2\operatorname{arccosh}\tfrac{11}{2}\simeq0.827$, and falls monotonically to $3/4$ in the pure-state limit. The first has $P/2T<1$ throughout, approaching unity as $\alpha\rightarrow\pi/3$ like $1+\log(3/4)/2\rho_1$, where the tree and the triangle degenerate onto the same three boundary points.

\bibliography{main}

\end{document}